\documentclass{article}

\usepackage[utf8]{inputenc} 
\usepackage[T1]{fontenc}    
\usepackage{hyperref}       
\usepackage{url}            
\usepackage{booktabs}       
\usepackage{amsfonts}       
\usepackage{nicefrac}       
\usepackage{microtype}      
\usepackage{xcolor}         

\usepackage{graphicx}
\usepackage{amsmath,amssymb}
\usepackage{bbm}
\usepackage{amsthm}

\usepackage[square,numbers]{natbib}
\usepackage[preprint]{neurips_2026}

\title{Steerable Single-step Retrosynthesis with Electron-pushing Discrete Flow Matching and Feynman-Kac Steering}

\author{%
  Octavian Susanu\textsuperscript{1*},
  Xuan-Vu Nguyen\textsuperscript{1*},
  Philippe Schwaller\textsuperscript{1,2} \\
  \textsuperscript{1}\'Ecole Polytechnique F\'ed\'erale de Lausanne (EPFL) \\
  \textsuperscript{2}National Centre of Competence in Research (NCCR) Catalysis \\
  \texttt{\{octavian.susanu,nguyen.nguyen,philippe.schwaller\}@epfl.ch}
}

\begin{document}

\maketitle

{\renewcommand{\thefootnote}{*}\footnotetext{Equal contribution.}}

\begin{abstract}
  Expert chemists perform retrosynthetic analysis in a strategic manner, where each reaction in a route is employed with a specific purpose, combined with mechanistic reasoning represented as electron-pushing arrows to visualize how bonds are formed or broken via electron redistribution.
  Yet contemporary single-step retrosynthesis models exhibit neither of these properties: their learned reaction distribution is fixed to that of the training data, and reactions are predicted via de-novo generation of the precursors or molecular graph edits, omitting the electron redistribution that governs the chemical transformations.
  We introduce Verso MAELLE, the first electron-pushing-based single-step retrosynthetic model, representing disconnections as continuous-time Markov chains (CTMC) comprised of edits in the electron space, and is trainable using a discrete flow matching formulation.
  Being a generative model, Verso MAELLE is optionally steerable at inference time through the Feynman-Kac (FK) framework, using rewards that promote certain types of disconnections.
  We show that besides in-distribution learning, Verso MAELLE can be steered toward cyclization reactions which are scaffold-constructive but underrepresented in the training corpus, via a ring-breaking reward.
  As a proof of concept, we show how the electron-pushing reasoning, learned from patented data, combined with a domino incentivizing reward, could steer Verso MAELLE toward creative and strategic disconnections, demonstrated on the key domino sequence that was used to synthesize Punctaporonin U, a complex natural product of the sesquiterpene family. 
\end{abstract}

\section{Introduction}

One of the challenges in experimental validation of \textit{in-silico} generated materials
and drugs is their synthesizability~\citep{nippa2025accelerating,guo2025generative}.
Decomposing a complex organic molecule into readily available precursors
is one of the most challenging and intellectually demanding tasks
in chemistry~\citep{corey1991logic}, being simultaneously a science
as well as an art. Corey formalized this endeavour into what is now known
as \textit{retrosynthesis}~\citep{corey1967general,Corey_Cheng_LogicofChemicalSynthesis}.  
Efforts in automating retrosynthesis into Computer-Aided Synthesis Planning (CASP)
began more than 50 years ago~\citep{corey1969computer,corey1972computer}, when Corey
and Wipke started with hard-coded reaction rules. Modern, data-driven
CASP approaches~\citep{genheden2020aizynthfinder,maziarz2024chemist}
involve combining a search algorithm~\citep{segler2018planning,
schwaller2020predicting,chen2020retro} with a single-step policy trained on a reaction corpus~\citep{schwaller2019molecular,tu2022permutation,sacha2021molecule}.

While this approach works well on simple and small molecules, recent work by 
\citet{armstrong2026strategy} has shown that complex targets such as natural products require a paradigm that goes beyond exhaustive tree search, incorporating \textit{strategic planning}, where each reaction has a different purpose at different stages of the synthesis: strategic disconnections often involve breaking the central scaffold of the molecules, while supporting reactions focus on manipulating functional groups \cite{mikulak2020computational}.
However, single-step policy models are fixed to the reaction distribution they are trained on. Moreover, the training data is often derived from patents \cite{Lowe2017}, which are biased toward routine reactions while scaffold-constructive ones such as cyclizations are underrepresented \cite{thakkar2020ring}.
Therefore, a mechanism to steer the single-step policy models at inference time is needed to adapt to diverse strategic requirements of synthesis planning.

The second facet of strategic retrosynthesis among synthetic chemists is the practice of reaction reasoning by drawing \textit{arrow-pushing} diagrams (also known as \textit{electron-pushing} or \textit{curved-arrow} diagrams) \cite{Corey_Cheng_LogicofChemicalSynthesis,clayden2012organic}.
The arrows describe a symbolic, directional transfer of electrons from a source to a sink.
The direction of the arrow cannot be drawn arbitrarily: electrons have to flow from an electron-rich donor, called a \textit{nucleophile}, to an electron-deficient acceptor, called an \textit{electrophile}.
The pattern of nucleophile and electrophile serves as the basis of organic synthesis and generalizes across diverse reaction types.
Reasoning about reactions through electron flow also tightly connects to another facet of strategic retrosynthesis that is employed by expert chemists: planning a reaction at the \textit{synthon} level before completing the functional groups \cite{Corey_Cheng_LogicofChemicalSynthesis,roh2026higher}.
Synthons are an abstract representation of the precursors, formed by a broad disconnection strategy, where hypothetical charges are placed on atoms to be later completed with compatible functional groups.

Previously, we developed MAELLE \cite{maelle}, an electron-pushing-based discrete flow matching reaction outcome prediction model.
MAELLE can be trained from large-scale end-to-end reactions, inferring the outcome of a reaction by simulating a continuous-time Markov chain (CTMC) acting on electrons using a vocabulary of arrow-pushing-inspired electron moves.
It was shown that the direction of the arrow-pushing moves produced by MAELLE is highly aligned with that of expert-annotated mechanistic steps, suggesting MAELLE can capture the pattern of nucleophile-electrophile in a reaction.
This makes MAELLE attractive to be used as our proposed strategically steerable single-step policy: its generative matching framing enables steering at inference using frameworks such as Feynman-Kac \cite{singhal2025general}, and its rollout trajectory resembles the arrow-pushing notation by design.

\begin{figure}[t]
    \centering
    \includegraphics[width=\linewidth]{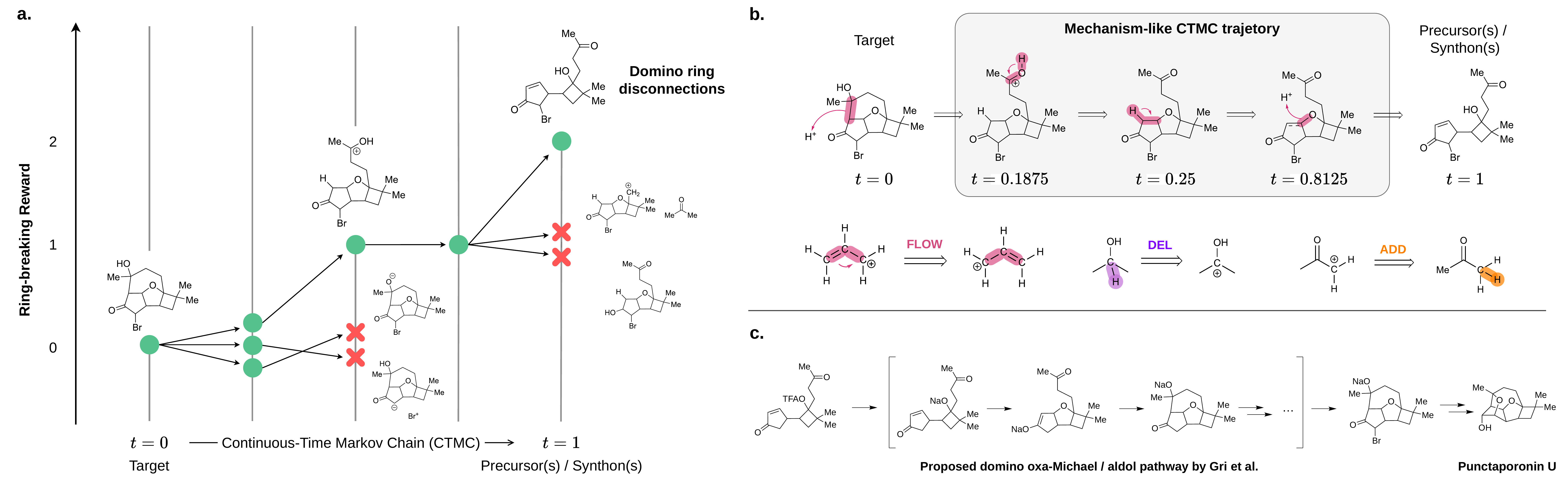}
    \caption{\textbf{Overview of FK-steered Verso MAELLE for single-step retrosynthesis},
demonstrated on a key intermediate in the total synthesis of Punctaporonin~U.
\textbf{a.} At test time, a reward steers a generative single-step retrosynthesis
model - Verso MAELLE in our case - toward constructive disconnections, here by
rewarding the number of rings broken between product and synthons.
\textbf{b.} Verso MAELLE as the retrosynthesis policy: the discrete states of the
CTMC are traversed by a set of operations acting on electrons - FLOW, DEL, and
ADD - yielding a pseudo-mechanistic, arrow-pushing-like disconnection.
\textbf{c.} The domino oxa-Michael/aldol sequence performed in the forward
direction in the literature \cite{gri2025concise} to synthesize the key precursor. FK-steered Verso
MAELLE captures this disconnection within its CTMC trajectories.}
    \label{fig:main}
\end{figure}

In this work, we introduce Verso MAELLE (\textit{verso} as in the back of a page), adapting MAELLE to the task of single-step retrosynthesis. Similar to MAELLE, Verso MAELLE employs a vocabulary of electron moves inspired by the arrow-pushing notations (Figure~\ref{fig:main}b). 
Since the moves only act on electrons, Verso MAELLE conserves heavy atoms by design and thus can plan retrosynthesis at the synthon level, focusing on the disconnection strategy, and the completion of synthons to precursors is outsourced to a simpler Transformer-based model.
We train Verso MAELLE on a single training corpus based on USPTO-480K \cite{Lowe2017,jin2017predicting}), and at inference time, the behavior of the model can be altered with Feynman-Kac steering and a reward that encodes synthesis strategy, promoting the construction of the main scaffold (Figure~\ref{fig:main}a).

We summarize our contributions as follows: (i) we introduce Verso MAELLE, the first single-step policy model that uses arrow-pushing on electrons to perform retrosynthetic disconnections, mimicking expert chemists' reasoning; (ii) given a fixed trained model, we demonstrate how the suggested precursors can be steered with a ring-breaking reward, which prompts the model to perform cyclization reactions that are underrepresented in the training data, without compromising reaction plausibility;
(iii) As a proof of concept, we show how this setup can generalize to mechanism-grounded novel reactions, demonstrated on a key and novel synthetic step that involves a domino sequence in the synthesis of Punctaporonin U \cite{gri2025concise}, forming two rings in one pot.

\section{Related work}
\subsection{Symbolic single-step retrosynthesis}
Most template-free retrosynthetic models cast the problem as a direct mapping from a product to its precursors, learned end-to-end over string or graph representations. Translation-based models treat the reaction as sequence-to-sequence transduction: the Molecular Transformer \cite{schwaller2019molecular}
generates precursor SMILES autoregressively from the product SMILES, and Graph2SMILES \cite{tu2022permutation} encodes the product as a 2D graph — recovering permutation invariance over the input — while still decoding the precursors as a SMILES string. Graph-edit models instead express the transformation as edits to the product graph: MEGAN \cite{sacha2021molecule} predicts an autoregressive sequence of atom- and bond-level edit operations that convert the product into its precursors.

Effective as they are on routine, patent-scale chemistry, these models operate over tokens or graph topology, not the electrons that mediate a disconnection, so a chemically sound disconnection is indistinguishable from a hallucinated one at the level of the representation. Additionally, they expose no distribution that could be tilted toward a desired outcome at inference time, redirectable only by retraining or re-weighting the data \cite{thakkar2020ring}.

\subsection{Generative approaches for single-step retrosynthesis}

Generative approaches to retrosynthesis aim to model the coupled distribution between products and reactants, learning to sample precursors conditioned on a target rather than scoring a fixed template library. Most of these methods reason at the level of molecular topology: a disconnection is cast as a transformation of the product graph over its nodes and edges — adding, deleting, or relabeling atoms and bonds — that converts the target into its precursors.

RetroBridge \cite{igashov2024retrobridge} frames retrosynthesis as a stochastic mapping between the two dependent distributions of products and reactants, learning a Markov bridge between the coupled endpoints. Generation proceeds by iteratively resampling categorical node and edge types along the bridge, so the transformation is expressed entirely as substitutions of atom and bond labels. Because the model operates on graph types rather than on the electrons that mediate the reaction, the disconnection it produces is a topological rewrite: it can insert or remove nodes and edges, but an arrow-pushing pathway is not guaranteed to be found between the two endpoints.

RetroProdFlow \cite{yadav2026retro} instead adopts discrete flow matching for end-to-end retrosynthesis, mapping products to reactants, and applies Feynman–Kac steering to the full generative process. However, it steers exclusively toward a roundtrip-accuracy reward. Such a reward only rewards recovering precursors that reconstruct the target, and does not incentivize any particular type of disconnection; in particular, it gives no preference to constructive reactions such as ring-forming steps, which are central to how chemists organize a synthetic strategy but underrepresented in patent reaction data. The steering therefore improves reconstructive validity without expressing the strategic choices that make a disconnection useful on complex targets.

Verso MAELLE addresses both of these gaps. Building on MAELLE \cite{maelle}, we model the retrosynthetic disconnection as a sequence of actions on electrons — FLOW, ADD, and DEL over the electron-occupation vector — rather than as node-and-edge edits on the molecular graph, so that each proposed disconnection carries an explicit electron-pushing mechanism. Having the
pseudo-mechanism, we are able to steer this generative process with rewards that encode genuine chemical strategy — such as the ring-breaking reward of Section~\ref{sec:fk-steering} — rather than mere reconstructive validity, biasing the model toward the strategically meaningful but underrepresented disconnections that characterize complex targets.

\section{Verso MAELLE - Electron-pushing-based and steerable single-step retrosynthesis}

\subsection{MAELLE overview}

\label{sec:background-maelle}

MAELLE~\cite{maelle} is a discrete flow matching (DFM) model that performs
pseudo-mechanistic forward reaction prediction by parametrizing a CTMC over the electron space of a molecular graph. 
This section provides a high-level overview of the method.
For a formal background on DFM and MAELLE, see Appendix \ref{app:background}

MAELLE represents a molecule as a graph of \emph{electron sites} - the bonds
between heavy atoms, the lone pairs on each atom, and a pooled site for the
hydrogens attached to each heavy atom - and tracks how many electron pairs
occupy each site. A reaction is then a redistribution of these electron pairs
across the sites, and MAELLE carries it out through three chemically
interpretable edits: \textsc{FLOW} moves an electron pair from one site to
another, forming or breaking a bond in the direction it is drawn (a source and a
sink, as in an arrow-pushing step); \textsc{ADD} introduces a new pair at a
site; and \textsc{DEL} removes one. Because the edits act on electrons rather
than on atoms, and hydrogens are supplied freely from their pooled site, the
transformation conserves every heavy atom by construction--the property
Verso~MAELLE relies on in Section~\ref{sec:synthon}. Sequentially applying these
edits turns the reactant into the product along a pseudo-mechanistic trajectory,
where the bond connectivity at any point is read off directly from which sites
currently hold electron pairs.

Given an end-to-end reaction, at training time, an offline process
recovers the set of electron edits connecting them by solving a balanced integer
optimal transport (OT) problem between the electron sites of the reactant and
those of the product, with a cost given by the shortest-path distance between
sites so that electrons are moved between nearby atoms. These edits define an
edit-based conditional mixture path between reactant and product, along which the
model - a graph neural network $u_\theta^t$ - is trained to predict, at any point
on the trajectory, the rate at which each remaining edit should fire
(Appendix~\ref{app:maelle}).

At inference only the reactant is known. MAELLE starts from the reactant graph
and evolves the trajectory in a fixed number of steps, at each step firing every
admissible electron edit independently and at random according to its predicted
rate. Because firing is stochastic, running multiple trajectories from the same
reactant yields a distribution over products; the frequency of each distinct
product defines the confidence used for top-$k$ ranking. This stochastic and
multi-trajectory sampling is what Verso~MAELLE later steers
(Section~\ref{sec:fk-steering}).

\subsection{Retrosynthesis over synthon space} \label{sec:synthon}

Retrosynthesis reasoning is not always atom-conservative: given
only the target product, a model predicts only the reactants, while the byproducts are typically omitted or predicted post-hoc.
Since MAELLE~\cite{maelle} operates only on electrons, it cannot take the byproducts into account and thus is unable to do end-to-end retrosynthesis prediction from products to reactants.

In order to leverage MAELLE's mechanistic reasoning for retrosynthesis prediction, we move our attention
toward \textit{synthons}. Synthons represent an abstraction
of the reactants, in which the core-molecular skeleton that will
be part of the target product is left untouched, but the leaving
groups are replaced by charges attributed to the core atoms
to which these were attached~\citep{corey1967general}. Such
charges should be representative of the
nucleophilic/electrophilic character of the reactive atoms~\citep{Corey_Cheng_LogicofChemicalSynthesis}.
As charges denote the presence or absence of electron pairs,
retrosynthetic prediction from target products toward synthons
is a heavy atom-conservative process, allowing therefore MAELLE
to be used for the task without modification.

Similarly to previous works~\citep{roh2026higher},
we define the charges added to the synthons based on the 
difference in Pauling electronegativity, $\chi$, between the synthon
atom and leaving group atom that are bonded in the reactant.
Therefore, when building a synthon from a reactant,
if $\chi_{\textit{i}} >\chi_{\textit{j}}$, the bond 
between atoms $\textit{i}$ and $\textit{j}$ is removed,
atom $\textit{i}$ receives the charge $-\textit{BO}$
and atom $\textit{j}$ receives the charge $+\textit{BO}$,
where both $\textit{i}$ and $\textit{j}$ represent heavy-atoms which
were bonded in the reactants, but out of which one is no longer
present in the target product and \textit{BO} represents the order
of the bond that was broken (bonds between identical atoms
are resolved as described in Appendix \ref{data_synthons}). Thanks to MAELLE's capability
of adding and removing hydrogen atoms at will, it is not necessary to
break such bonds with hydrogen during synthon formation. 
Constructing the synthons as described allows MAELLE to interpolate a pseudo-mechanism
from the products to the synthons through its optimal transport
algorithm.

We refer to this product-to-synthon adaptation of MAELLE as Verso MAELLE. 
Formally, we denote the molecular graph of the target product by $\mathcal{G}_0$ and that of the resulting synthons by $\mathcal{G}_1$. 
Verso MAELLE models the retrosynthetic transformation as a CTMC trajectory $(\mathcal{G}_t)_{t \in [0,1]}$, interpolating between the target product at $t=0$ and the synthons at $t=1$.

\subsection{Completing the synthons to reactants}

Previously, synthons have been completed using a constructed vocabulary of leaving groups,
casting the problem into a multi-class classification problem~\citep{somnath2021learning}.
In Appendix~\ref{app:synthon_completion_data}, we show that to each leaving group,
an \textit{enhanced SMILES} string can be associated, such that the SMILES of the synthon and
the enhanced SMILES of the leaving group can be combined into the SMILES of the reactant.
Therefore, the synthon completion problem can be treated as a neural
machine translation task, where a model learns to predict the enhanced
SMILES representation of the corresponding leaving group, given the SMILES
of a synthon, followed by completion to the reactant during post-processing.
Therefore, we train a \textit{Synthon Transformer} model, an implementation of
the Transformer architecture~\citep{schwaller2019molecular,vaswani2017attention} on the task of
synthon completion.

\subsection{Inference-time steering with Feynman-Kac}
\label{sec:fk-steering}

We aim to steer the learned conditional distribution $p_\theta(\mathcal{G}_1 | \mathcal{G}_0)$ toward strategically desirable disconnections by defining a reward-tilted target distribution. More generally, the reward can depend on the entire trajectory $\tau=(\mathcal{G}_0,\mathcal{G}_h,\ldots,\mathcal{G}_1)$, rather than only on its endpoints. We therefore define the reward-tilted distribution over trajectories as:

\begin{equation}
    p_{target}\left(\tau|\mathcal{G}_0\right) = \frac{1}{Z}p_{\theta}\left(\tau|\mathcal{G}_0\right)\textrm{exp}\left(\lambda r\left(\tau\right)\right)
    \label{target_distribution}
\end{equation}

where $\lambda$ controls the strength of the distribution tilting, $r(\tau)$ is a reward function that evaluates the retrosynthetic transformation along the trajectory $\tau$, and $Z$ is the normalization constant. The corresponding distribution over the final synthons is obtained by marginalizing over all trajectories that lead to $\mathcal{G}_1$:

\begin{equation}
    p_{target}\left(\mathcal{G}_1|\mathcal{G}_0\right) = \sum_{\tau:\tau_1=\mathcal{G}_1} p_{target}\left(\tau|\mathcal{G}_0\right)
\end{equation}

where $\tau_1$ denotes the final molecular graph of trajectory $\tau$. Endpoint-based rewards, such as the ring-breaking reward, are recovered as a special case when $r(\tau)=r(\mathcal{G}_1,\mathcal{G}_0)$.

To achieve sampling from the distribution in Equation \ref{target_distribution}, we use Feynman-Kac (FK) steering~\citep{singhal2025general}, an inference-time steering method that has seen application in discrete flow-matching retrosynthesis models~\citep{yadav2026retro}. FK steering guides the trajectory toward high-reward outcomes by iteratively tilting the transition kernels $p_\theta\left(\mathcal{G}_{t+h}|\mathcal{G}_t, \mathcal{G}_0\right)$ with a function, $G_t\left(\tau_{0:t}\right)$, known as \textit{potential}, where $h$ is the Euler timestep and $\tau_{0:t}=(\mathcal{G}_0,\mathcal{G}_h,\ldots,\mathcal{G}_t)$ denotes the partial trajectory up to time $t$. Thus, starting from an initial FK distribution $p_{FK,0} \propto p_{source}$, subsequent FK distributions are defined as:

\begin{equation}
    p_{FK, t+h}\left(\tau_{0:t+h} | \mathcal{G}_0\right) = \frac{1}{Z_{t+h}} p_{\theta}\left(\mathcal{G}_{t+h}|\mathcal{G}_{t}, \mathcal{G}_0\right)G_{t+h}\left(\tau_{0:t+h}\right) p_{\theta}\left(\tau_{0:t}|\mathcal{G}_0\right) \prod_{s=h}^{t} G_s\left(\tau_{0:s}\right),
    \label{eq:fk_distribution}
\end{equation}

where the normalization constant is $Z_{t}=\mathbb{E}_{p_\theta( \cdot | \mathcal{G}_0)}\left[\prod_{s=h}^{t}G_s\right]$ and the condition $\prod_{s=h}^1G_s\left(\tau_{0:s}\right) = \textrm{exp}\left(\lambda r\left(\tau\right)\right)$ is imposed. We note that, unlike the FK formulation in ~\citep{singhal2025general}, we do not tilt $p_{source}$ with the potential function because our trajectories start with the molecular graph $\mathcal{G}_0$ of the target, rather than a sample from a simple prior.

As direct sampling from $p_{FK, 1}$ is intractable, we apply sequential Monte Carlo (SMC) methods. SMC generates $k$ interacting flow processes, called \textit{particles}. The particles are resampled at each step of the trajectory or, alternatively, at a certain resampling interval. By choosing the SMC's proposal generator $q$ to be equal to the model's transition kernel $p_\theta\left(\mathcal{G}_{t+h}|\mathcal{G}_t, \mathcal{G}_0\right)$~\citep{singhal2025general}, the importance weights of the particles during resampling become $w_t^i = G_t\left(\tau_{0:t}^i\right)$, with $i=1,\ldots,k$.

The potential $G_t$ scores a trajectory by accumulating a reward $r_\phi$ of the particles. A popular accumulation scheme is using SUM potential~\cite{singhal2025general}:

\begin{equation}
    G_t\left(\tau_{0:t}\right) = \textrm{exp}\left(\lambda \sum_{s=h}^t r_{\phi}\left(\tau_{0:s}\right)\right),
    G_1\left(\tau\right) = \frac{\textrm{exp}\left(\lambda r\left(\tau\right)\right)}{\prod_{s=h}^{1-h}G_s\left(\tau_{0:s}\right)}
    \label{eq:sum_potential}
\end{equation}

Intermediate rewards can be defined by estimating the final reward from the partial trajectory. For a trajectory-dependent reward, a natural choice is the expected final reward conditioned on the partial trajectory, $r_\phi\left(\tau_{0:t}\right)=\mathbb{E}_{p_\theta}\left[r\left(\tau\right)|\tau_{0:t}\right]$. In our case, we can take advantage of the fact that the intermediate states along Verso MAELLE's trajectories are chemically interpretable molecular graphs and that the electron edits performed along the trajectory can be tracked explicitly. We therefore define the intermediate rewards by directly evaluating the same reward function on the partial trajectory: $r_\phi\left(\tau_{0:t}\right)=r\left(\tau_{0:t}\right)$.

\subsection{Chemically and strategically interpretable reward functions}

As the electron-pushing trajectories of Verso MAELLE result in chemically interpretable intermediate molecular graphs, it is possible to design rewards that encapsulate the strategic thinking of synthetic chemists.
Cyclic structures are common motifs in organic molecules, and expert organic chemists may choose
to organize their synthetic strategy around disconnecting such cyclic structures.
However, cyclization reactions are fairly rare in USPTO 480K (Table~\ref{tab:synthon_dataset_stats}).
Therefore, we define a \textit{ring-breaking reward} ($r_\text{ring}$), which
counts the number of rings that have been broken in the current molecular graph, prioritizing sampling particles with more rings broken:

\begin{equation}
    r_{\textrm{ring}}\left(\mathcal{G}_t, \mathcal{G}_0\right) = R_{\mathcal{G}_0} - R_{\mathcal{G}_t},
    \label{ring_breaking}
\end{equation}

where $R_{\mathcal{G}}$ counts the number of rings present in the molecular graph $\mathcal{G}$.
This reward is used for all USPTO 480K-related results.

Additionally, we also implement a \textit{domino reward} ($r_\text{domino})$.
In organic synthesis, \textit{domino} (sometimes called \textit{cascade}) reactions are processes where multiple bond-forming transformations occur under the same reaction condition, where subsequent steps are triggered by the previous ones \cite{tietze1996domino,nicolaou2009art,gri2025concise}.
Domino reactions are of significant interest in synthetic chemistry due to their conciseness and the rapid construction of scaffold complexity.
Reasoning about domino retrosynthetic disconnections involves visualization of multiple electron-pushing moves.
To promote scaffold-constructive domino disconnections, we define the reward as follows:

\begin{equation}
     r_\textrm{domino}\left(\mathcal{G}_t, \mathcal{G}_{t-h},..., \mathcal{G}_0\right) = 
     c_{\textrm{ring}}r_{\textrm{ring}}\left(\mathcal{G}_t, \mathcal{G}_0\right) + 
     c_{\textrm{flow}}r_{\textrm{flow}}\left(\mathcal{G}_t, \mathcal{G}_{t-h},..., \mathcal{G}_0\right) +
     c_{\textrm{charge}}r_{\textrm{charge}}\left(\mathcal{G}_t\right),
    \label{eq:combined}
\end{equation}

where $r_\text{flow}$ rewards the model for performing multiple electron-pushing moves, measured as the number of \textsc{FLOW} moves performed across the whole trajectory up to particle $\mathcal{G}_t$; $r_\text{charge}$ penalizes the molecular graphs with many charges:
    $r_{\textrm{charge}}(\mathcal{G}) = -\sum_{i=1}^{N_{\textrm{atoms}\;\mathcal{G}}}\left|
    C_i\right|$,
and thus prevent the model from exploiting $r_\text{flow}$ by breaking unrelated bonds; $c_\textrm{ring}$, $c_\textrm{flow}$, and $c_\textrm{charge}$ are corresponding coefficients of the reward terms.
Due to the lack of a large-scale domino reaction dataset, in the scope of this work, we only demonstrate this reward qualitatively on a key domino reaction in the total synthesis of Punctaporonin U \cite{gri2025concise} (Figure \ref{fig:main}c).
We also emphasize that the definitions of $r_\textrm{flow}$ and $r_\textrm{charge}$ are only feasible thanks to the chemically interpretable CTMC trajectories of Verso MAELLE, in contrast to other categorical graph-based generative methods.

\section{Results}

\subsection{Experiments}

We pre-process the USPTO 480K dataset~\citep{Lowe2017} to extract the
synthons
and leaving groups, as described in Appendix~\ref{data_synthons},
obtaining the Synthon
USPTO 480K dataset. Using this dataset, we train Verso MAELLE on the product
to synthon prediction task, obtaining the base Verso MAELLE model.
Apart from vanilla inference, we also experiment with FK steering, the results of which are referred to as FK Verso MAELLE.
We perform particle resampling at a set of timesteps $\mathcal{R}$ that includes the final timestep $t=1$. At timesteps $t\notin\mathcal{R}$, we set $G_t=1$. At resampling steps $t\in\mathcal{R}$, we use the SUM potential defined in Equation~\ref{eq:sum_potential}.
As for the reward, we choose the ring-breaking reward defined in 
Equation~\ref{ring_breaking}. 
As baselines, we train a Molecular Transformer (MT) \cite{schwaller2019molecular} and Graph2SMILES (G2S)~\citep{tu2022permutation} to predict the synthons.
Details on the training and
inference configurations used are found in Appendix~\ref{training_details}.

Synthons predicted by all models are converted to canonical SMILES, and then completed to reactant molecules using Synthon Transformer, trained to translate the SMILES string of the synthons to that of the functional groups, enhanced with their attachment points. For all metrics that require completed reactant molecules, we take the top-1 completion of Synthon Transformer for each synthon prediction as its corresponding reactants. 
Further details on synthon completion data format and Synthon Transformer's performance can be found in Appendix \ref{app:synthon_transformer}.

To quantify in-distribution learning, we use \textit{top-k synthon exact match accuracy}, measuring how frequently the ground-truth synthons are found among the top-k predictions.
Since USPTO-480K is split randomly among training, validation, and test subsets, top-k synthon accuracy quantifies in-distribution learning of models.
In terms of reaction plausibility, we measure \textit{top-k roundtrip coverage} and \textit{accuracy}, using Molecular Transformer \cite{schwaller2019molecular} acting on reactants completed from synthons as described above. 
We use the Molecular Transformer model trained on the mixed version of USPTO
480K, whose weights are downloaded from \url{https://github.com/pschwllr/MolecularTransformer}.

To further quantify the effect of FK steering towards ring-breaking disconnections, we 
measure \textit{top-k
ring-breaking} coverage, which checks if a disconnection involving a ring-breaking
is found within the first \textit{k} predictions of the model. 
At the candidate level, we report \textit{average cumulative number of rings broken}, which computes the sum of the number of rings broken in the top-$k$ predictions and then averages over the entire prediction population up to rank $k$.

\subsection{Verso MAELLE is capable of in-distribution learning}

In this section, we address how well Verso MAELLE learns from its training data, using the popular Molecular Transformers (MT) and Graph2SMILES (G2S) as baselines.
The top-\textit{k} synthon and roundtrip accuracies of the models
over USPTO 480K Synthon are presented in Table~\ref{tab:topk-uspto}.
Although at lower ranks, the top-$k$ synthon accuracy of Verso
MAELLE is behind that of MT and G2S (0.5221 \textit{top-1} for Verso MAELLE whereas G2S achieves 0.6840), the
difference becomes less significant for later ranks and at top-20, Verso MAELLE achieves a slightly better performance
than G2S by 1.02 pp, while trailing behind MT by 5.86 pp. 
The top-k synthon accuracy of Verso MAELLE increases quickly with more samples: the difference between Verso MAELLE and G2S changes from -16.19 pp to +1.02 pp, implying the reaction space that Verso MAELLE produces is more diverse.

Top-k exact match only measures how well a model captures its training data.
However, in the context of retrosynthesis, multiple reactions could be used to synthesize the same target, and thus the quality of the reactions proposed is more important.
We show that Verso MAELLE can propose chemically valid disconnections
through the top-$k$ roundtrip coverage and accuracy.
These metrics show a smaller performance gap between models.
While at top-1 Verso MAELLE's roundtrip coverage is slightly behind that of MT and G2S
by a maximum of 2.93 pp, Verso MAELLE consistently becomes the second best performing
model for this metric, being ahead of G2S by 2.48 pp at top-20 and achieving a
similar performance to MT, with a difference of 0.12 pp.

As opposed to roundtrip coverage, which measures whether at least one prediction at top-$k$ yields
the correct product in the forward direction, the roundtrip accuracy measures, across all suggestions, how many of them yield
the desired product. 
A good single-step retrosynthesis model needs to predict not just one but several chemically valid reactions.
In this regard, Verso MAELLE is comparable with MT, while substantially outperforming G2S from $k=3$ onward.
We note that roundtrip accuracy serves as a proxy for chemical plausibility rather than an absolute metric, since the forward prediction model lacks the context of reaction conditions, which have a heavy influence on the outcome of a reaction but are not output by retrosynthesis models.

\begin{table*}[t]
\centering
\caption{Top-\textit{k} synthon accuracy, followed by top-\textit{k} coverage and top-\textit{k} roundtrip accuracy of the completed synthons,
all on USPTO 480K Synthon.}
\small
\setlength{\tabcolsep}{3pt}
\label{tab:topk-uspto}
\resizebox{\textwidth}{!}{%
\begin{tabular}{@{}lccccc|ccccc|ccccc@{}}
\toprule
& \multicolumn{5}{c}{\textbf{Top-\textit{k} Synthon Accuracy}} & \multicolumn{5}{c}{\textbf{Top-\textit{k} Roundtrip Coverage}} & \multicolumn{5}{c}{\textbf{Top-\textit{k} Roundtrip Accuracy}} \\
\cmidrule(lr){2-6} \cmidrule(lr){7-11} \cmidrule(lr){12-16}
\textbf{Model} & $k=1$ & $k=3$ & $k=5$ & $k=10$ & $k=20$ & $k=1$ & $k=3$ & $k=5$ & $k=10$ & $k=20$ & $k=1$ & $k=3$ & $k=5$ & $k=10$ & $k=20$ \\
\midrule
MT                       & \underline{0.6465} & \textbf{0.8183} & \textbf{0.8642} & \textbf{0.9016} & \textbf{0.9215} & \underline{0.8515} & \textbf{0.9586} & \textbf{0.9753} & \textbf{0.9871} & \textbf{0.9914} & \underline{0.8515} & \underline{0.6532} & \underline{0.5436} & \underline{0.4033} & 0.3062 \\
G2S                      & \textbf{0.6840} & \underline{0.8052} & \underline{0.8286} & \underline{0.8451} & 0.8527 & \textbf{0.8631} & 0.9381 & 0.9512 & 0.9614 & 0.9651 & \textbf{0.8631} & 0.5900 & 0.4660 & 0.3354 & 0.2652 \\
\textbf{Verso MAELLE}    & 0.5221 & 0.7308 & 0.7915 & 0.8365 & \underline{0.8629} & 0.8338 & \underline{0.9488} & \underline{0.9721} & \underline{0.9850} & \underline{0.9902} & 0.8338 & \textbf{0.6598} & \textbf{0.5498} & \textbf{0.4040} & \underline{0.3363} \\
\textbf{FK Verso MAELLE} & 0.4936 & 0.6886 & 0.7398 & 0.7909 & 0.8197 & 0.8008 & 0.9223 & 0.9476 & 0.9690 & 0.9784 & 0.8008 & 0.6033 & 0.4899 & 0.3872 & \textbf{0.3596} \\
\bottomrule
\end{tabular}%
}
\vspace{4pt}
\footnotesize{The FK steered model uses the \textit{ring-breaking reward}, with 4 particles and $\lambda$ = 1.}
\end{table*}

\begin{figure}[t]
    \centering
    \includegraphics[width=0.8\linewidth]{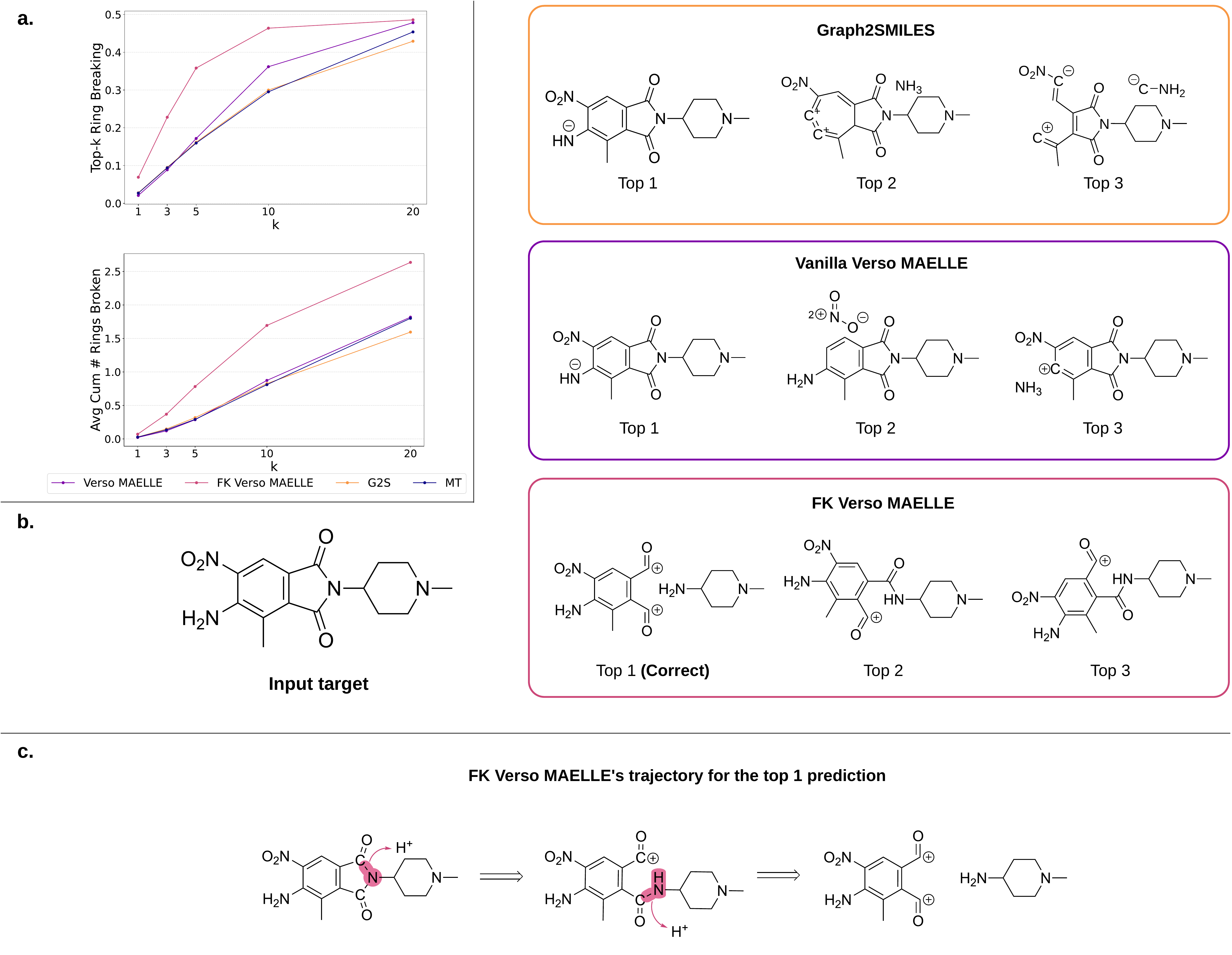}
    \caption{Ring breaking comparison between the models over USPTO 480K Synthon.
    \textbf{a.} Ring breaking metrics over the USPTO 480K Synthon
    dataset. The FK-steered Verso MAELLE model recovers more
    ring-breaking disconnections and cumulatively breaks on average
    more rings than both non-steered baseline models for equal
    $k$. \textbf{b.} An imide, whose ground truth disconnection is a double amide bond cleavage. FK steered Verso MAELLE recovers only ring-breaking
    disconnections up to \textit{top-3}, including the ground truth
    synthon (highlighted), while no other model recovers chemically 
    plausible ring-breaking
    disconnections for the given product. \textbf{c.} Mechanistic trajectory
    generated by FK Verso MAELLE between product and ground truth predicted
    synthon.}
    \label{fig:ring_breaking_final}
\end{figure}

\subsection{Verso MAELLE is steerable towards rare disconnections}

From the base Verso MAELLE, we run another round of inference on the test set of USPTO 480 Synthon data, but the model's trajectories are steered with the ring-breaking reward (Eq~\ref{ring_breaking}).
It is expected that since the model's prediction is biased toward an objective, its top-k exact-match would decrease.
Indeed, its synthon
accuracy is trailing behind that of
vanilla Verso MAELLE and other models, but still in the reasonable range: the accuracy is at 49.36\% and 81.97\% at $k=1$ and $k=20$, respectively.
Furthermore, FK steering does not prevent the model from recovering good reactions, characterized by the roundtrip coverage.
While it is still lower than vanilla Verso MAELLE and MT, the roundtrip coverage is comparable to that of G2S, outperforming the latter from top-10 onward.
The same trend is observed for roundtrip accuracy at $k=3$, $5$, and $10$, and at $k=20$, the FK-steered model has the highest accuracy among the four, at 35.96\%.
This is explained by the fact that the FK-steered Verso MAELLE proposes 12 predictions on average, compared with 17 for the other models.
The lower number of predictions under FK steering could be attributed to the particle resampling process (Eq~\ref{eq:fk_distribution}) that collapses diverse particles into a few that have high reward.

The effectiveness of the steering
is measured through the metrics \textit{top-k ring breaking} and \textit{average
cumulative number of rings broken}, presented in Figure~\ref{fig:ring_breaking_final}a.
We observe from the \textit{top-k} ring breaking that the FK
steered model prioritizes, compared to the other models, at each rank
disconnections that break a ring. 
The low frequency of ring-breaking disconnections in the training data (Table~\ref{tab:synthon_dataset_stats}) is reflected by the low top-1 ring-breaking of the non-steered models at roughly 3\%.
Under the influence of the ring-breaking reward, the FK-steered Verso MAELLE is more likely to perform retro cyclizations at $k=1$, $3$, $5$, and $10$.
At top-20, however, both vanilla and FK-steered Verso MAELLE are equally probable to have at least one ring-breaking reaction in their predictions, at around 48\% of the time, suggesting that with enough samples, a ring-breaking disconnection would eventually be recovered, but FK-steering could help surface these disconnections with fewer examples.
Besides being able to perform cyclization at lower sampling, FK-steered Verso MAELLE also breaks rings in different ways, as proved by the cumulative number of rings broken: 
at 20 different samples, the steered Verso MAELLE breaks 2.63 rings on average, compared with 1.81 of the vanilla inference.
Furthermore, we illustrate the higher affinity
for ring-breaking reactions of the FK-steered Verso MAELLE model with
a qualitative example in Figure~\ref{fig:ring_breaking_final}b,
where all \textit{top-3} disconnections of the steered model are
ring-breaking, whereas the non-steered models recover no chemically plausible ring-breaking disconnections. 
Notably, the top-3 prediction of G2S resembles a hallucinated disconnection of the benzene ring, while the transformation suggested by Verso MAELLE is grounded in the electron-pushing notation, as shown in Figure~\ref{fig:ring_breaking_final}c.

Together, these results highlight the strength of Verso MAELLE combined with the FK framework. The model does not have to be retrained to adapt to a new objective and thus maintains the chemistry learned during training, as reflected by exact-match and roundtrip metrics. 
On the other hand, guided by an objective function, it can recover more ring-breaking disconnections, overcoming the shortage of this transformation type in the training data.

\subsection{Domino reactions: A case study of Punctaporonin U}

To assess whether steering through mechanistically grounded rewards grants Verso MAELLE 
the potential to recover out-of-distribution disconnections, we test
its performance on a key step in the total synthesis of Punctaporonin U, a complex natural product of the sesquiterpene family,
featuring a pentacyclic, cage-like structure, that was synthesized successfully
by~\citet{gri2025concise} (Figure~\ref{fig:main}c). The key step involves a one-pot construction of the tetracyclic
core via a domino sequence oxa-Michael/aldol/bromination (Figure~\ref{fig:main}c). 
Such a domino sequence is highly unlikely to be available in the training corpus, but we envision that from the electron-pushing moves learned from the training data combined with a reward that incentivizes domino reactions (Eq~\ref{eq:combined}), Verso MAELLE could recover this disconnection, as the proof of concept for the flexibility of the method.
Hyperparameters for the rewards and FK can be found in Appendix~\ref{training_details}.

First, it is noteworthy to mention that, by analyzing
the mechanism of the transformation in the original publication~\citep{gri2025concise}, we observe
that only the oxa-Michael/aldol part of the sequence complies with our definition
of domino reaction: a reaction where each step is causally linked to the
previous one (see Section~\ref{sec:fk-steering}). The bromination is achieved
through the use of a base that forms the enolate; this enolate formation
does not have a causal relationship with the previous steps of the sequence.
Therefore, we consider that a model succeeds if it manages to recover
the retro-aldol/retro-oxa-Michael (retro equivalent of the forward sequence
oxa-Michael/aldol) part of the disconnection.

\begin{figure}[t]
    \centering
    \includegraphics[width=0.8\linewidth]{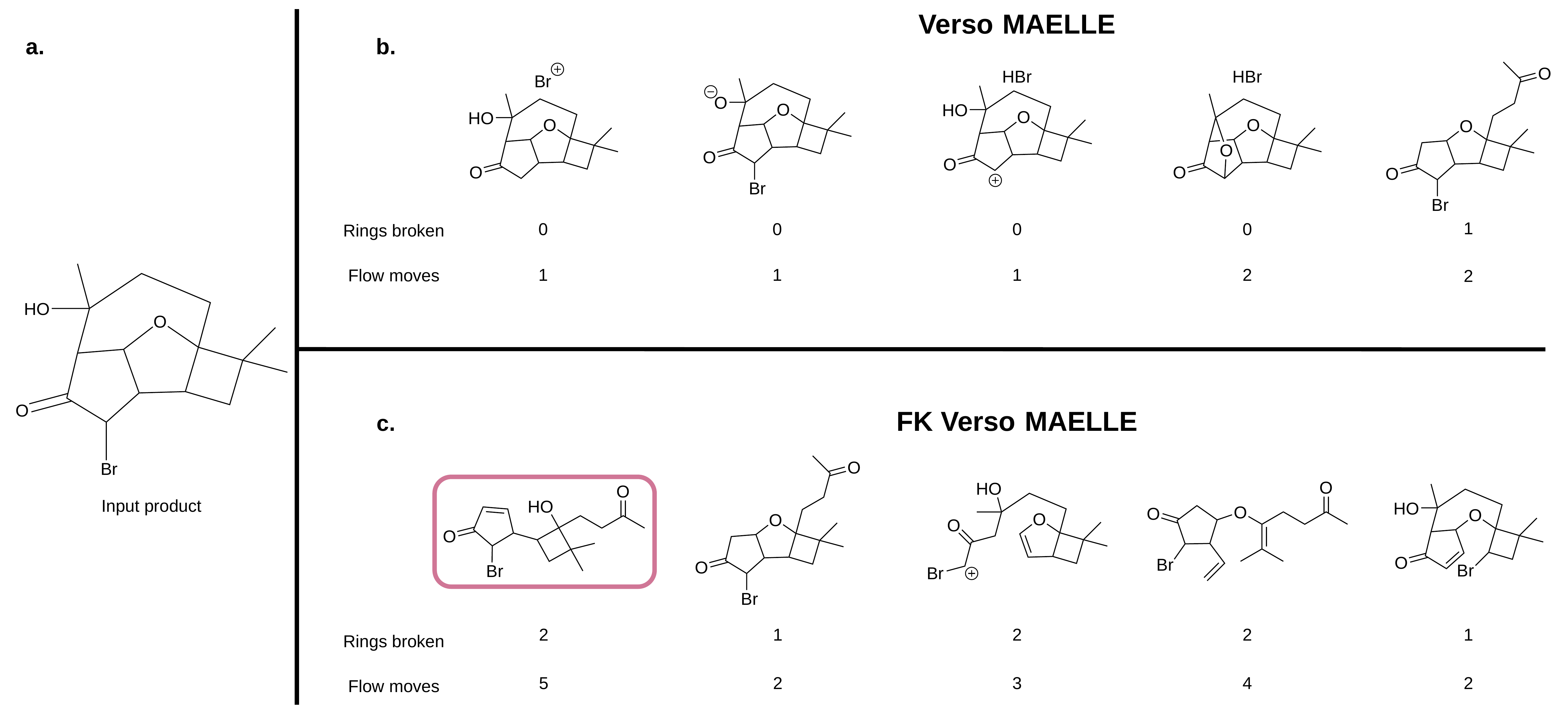}
    \caption{Top-5 synthon predictions of Verso MAELLE and FK Verso MAELLE steered with the
    \textit{domino} reward (see Equation~\ref{eq:combined}) for the key intermediate
    in Punctaporonin U's synthesis by~\citet{gri2025concise}. The results
    presented are for a given seed, and are accompanied
    by the number of ring broken and maximum number of flow moves
    across the different samples by which the synthons are reached.
    We consider a prediction
    successful if it recovers the causally related steps of the ground truth disconnection.
    \textbf{a. } Input to the models. \textbf{b. } In the absence of steering, Verso MAELLE
    is not incentivized to promote either ring-breaking disconnections or domino reactions.
    \textbf{c .} Steered FK Verso MAELLE manages to recover the ground truth retro-aldol/retro-oxa-Michael, a domino ring breaking disconnection. Moreover, all of the top-5 steered model's disconnections break rings, with the first, third, and fifth predictions being domino disconnections,
    while the fourth one involves two, non-causally related disconnections.}
    \label{fig:punctaporonin}
\end{figure}

We present the \textit{top-5} synthon predictions of Verso MAELLE and of FK Verso MAELLE
steered with the domino reward in Figure~\ref{fig:punctaporonin}.
While all the disconnections are plausible from a synthetic point of view, the non-steered Verso MAELLE performs neither ring-opening nor domino disconnections. 
The \textit{top-1} and \textit{top-3} correspond to a ketone
alpha-bromination, the \textit{top-2} represents a retro-protection of an alcohol and
the \textit{top-4} represents a retro $\textrm{S}_{\textrm{N}}2$ disconnection, which actually recovers
Punctaporonin U. The only exception is the last prediction, which manages to break a 7-membered ring
through a retro-aldol disconnection, one of the disconnections involved in the ground truth one.
The model, however, is not incentivized in any way to continue
the disconnection in order to recover the ground truth.

On the other hand, the FK-steered Verso MAELLE model manages to recover as its \textit{top-1}
prediction the ground-truth domino retro-aldol/retro-oxa-Michael disconnection
(Figure~\ref{fig:punctaporonin}c). We note that by averaging over 6 different seeds,
the ground truth synthon is found at rank 6 on average. In Figure~\ref{fig:main}b we depict one of the
CTMC trajectories by which the model reaches the desired synthon. We observe that
this pseudo-mechanism closely mirrors, in the opposite
direction, the mechanism proposed by~\citet{gri2025concise}, depicted in
Figure~\ref{fig:main}c: the first two steps represent a retro-aldol
reaction, while the next two denote a retro-oxa-Michael addition, thus highlighting the interpretability of the trajectories produced by the model. It is thanks
to this interpretability that we are able to employ rewards that encode
chemical strategies, such as the one that we used for this case study. This
domino reward allows us to recover also other disconnections that can be classified
as ring-breaking domino disconnections, with their steps causally linked to one another. The
prediction at \textit{top-3} reaches the product in the forward direction by attacking the
carbocation with the enol-ether functional group of the dihydrofuran moiety, followed by subsequent
trapping of the resulting oxonium cation by the alpha-position of the ketone. The prediction at
\textit{top-5} can be interpreted as a Michael addition, achieved by converting the synthon into a Gilman
reagent, followed by trapping of the resulting enolate with a bromine cation. Even though its steps
are not causally related, the prediction at \textit{top-4} is composed of a [2+2] photocycloaddition
and an aldol condensation. Thus, while we do not assess the experimental feasibility of these discussed
other proposals, we emphasize that steering through rewards aligned with the strategies
developed by expert synthetic chemists grants
FK-steered Verso MAELLE the potential to recover diverse, creative, out-of-distribution disconnections,
including the ground-truth one for Punctaporonin U.

\section{Conclusion}

We introduced Verso MAELLE, the first electron-pushing-based single-step retrosynthesis model, which combines discrete flow matching with Feynman-Kac inference-time steering to promote strategically meaningful disconnections. Verso MAELLE models retrosynthetic transformations as continuous-time Markov chain trajectories of electron edits, converting target products into synthons through interpretable, pseudo-mechanistic pathways. The resulting synthons are subsequently completed into precursors using a separate Synthon Transformer.

We demonstrated that Verso MAELLE learns the reaction distribution of the training dataset while allowing its predictions to be steered toward underrepresented reaction types without retraining. In particular, FK steering with a ring-breaking reward promotes scaffold-constructive disconnections while retaining competitive roundtrip performance. As a proof of concept beyond the training distribution, we further showed that a more complex reward can guide Verso MAELLE toward domino ring-breaking disconnections, including the key retro-aldol/retro-oxa-Michael sequence in the total synthesis of Punctaporonin U. The recovery of this literature-reported disconnection illustrates the potential of mechanism-level steering to uncover strategically relevant transformations that are rare or absent in patent-derived reaction data.

\paragraph{Limitations and future work}
While the Punctaporonin U case study demonstrates the potential of Verso MAELLE for out-of-distribution retrosynthesis, a systematic evaluation on a broader set of complex targets outside the USPTO dataset is needed to establish the reliability and generality of mechanism-level steering. Developing a benchmark of such strategic disconnections is therefore an important direction for future work.

Another direction is to expand the range of chemically meaningful rewards. The ring-breaking and electron-flow rewards considered here capture specific aspects of synthetic strategy, but many principles underlying effective synthesis planning remain difficult to quantify. Encoding these principles as FK rewards could enable more flexible, chemist-aligned retrosynthetic predictions.

Verso MAELLE also has several methodological limitations. Its trajectories are pseudo-mechanistic: they represent electron-redistribution pathways derived from an optimal-transport formulation rather than physically validated sequences of elementary reaction steps. Although these trajectories provide an interpretable representation for steering disconnections, they should not be interpreted as experimentally established reaction mechanisms. Furthermore, the current model does not account for stereochemistry. Extending the edit vocabulary to represent stereocenters and bond geometry would broaden its applicability to stereochemically complex targets.

Finally, integrating Verso MAELLE and the Synthon Transformer into multistep retrosynthesis search algorithms \cite{segler2018planning,chen2020retro} would provide a natural next step toward strategy-guided synthesis planning for complex molecules.

\section*{Acknowledgement}
X.V.N acknowledges the support from the Marie Skłodowska-Curie Actions Doctoral Networks grant agreement “Explainable AI for Molecules - AiChemist” no. 101120466.
P.S. acknowledges the support from the NCCR Catalysis (grant number 225147).


\bibliography{citation}


\appendix

\section{Background} \label{app:background}
\subsection{Discrete Flow Matching} \label{app:dfm}

Discrete flow matching (DFM)~\citep{campbell2024generative, gat2024discrete} is a paradigm
in which a generative model parametrizes a Continuous Time Markov Chain (CTMC) that transports
a source distribution $p_0$ at $t=0$ to a target distribution $p_1$ at
$t=1$. The interpolation between the two is done through a marginal probability
path $p_t$. As sampling directly from the marginal is intractable, a conditional
path, $p_t\left(x|x_0, x_1\right)$, known as the \textit{conditional mixture path}
is employed, where $x_0$ and $x_1$ are drawn from the coupling $\pi
\left(x_0, x_1\right)$ between the source and target distributions. The conditional
path must satisfy boundary conditions $p_0(x\mid x_0,x_1)=\delta_{x_0}(x)$ and
$p_1(x\mid x_0,x_1)=\delta_{x_1}(x)$.

DFM can be applied in the context of modifying a sequence of tokens
to another sequence of equal length. Let us therefore consider $\mathcal{T}$
a vocabulary and $\mathcal{X}=\mathcal{T}^{*}=\bigcup_{n=0}^{N}\mathcal{T}^{n}$
the space of token sequences of length at most $N$. For two sequences of equal
length $n$, the conditional mixture path chosen is the factorized,
token-wise mixture path~\citep{gat2024discrete}:

\begin{equation}
  p_t(x^i \mid x_0, x_1)
  = (1-\kappa(t))\,\delta_{x_0^i}(x^i) + \kappa(t)\,\delta_{x_1^i}(x^i),
  \qquad
  p_t(x \mid x_0, x_1) = \prod_{i=1}^{n} p_t(x^i \mid x_0, x_1),
  \label{eq:tokenwise}
\end{equation}

where the schedule $\kappa:[0,1]\to[0,1]$ is monotone with $\kappa(0)=0$,
$\kappa(1)=1$. The CTMC which generates the probability path described in
Equation~\ref{eq:tokenwise} is characterized by the following token-wise,
marginal rate matrix:

\begin{equation}
  u_t(x^i \mid x_t)
  = \frac{\kappa'(t)}{1-\kappa(t)}\,
    \bigl[\,p_{1\mid t}(x^i \mid x_t) - \delta_{x_t^i}(x^i)\,\bigr],
  \label{eq:tokenwise-rate}
\end{equation}

where $p_{1\mid t}(x^i \mid x_t)$ is the posterior over the target
token at position $i$. Therefore, training reduces to learning
the posterior per token position using a cross-entropy objective.

The limitation of interpolating only between sequences of the same length
has been overcome by~\citet{havasi2025edit}, which introduces a set of
\textit{edit operations}, namely deletion, insertion and substitution
that allow interconversion between strings of different length.
In this setting, the problem is lifted to an auxiliary space
of aligned, padded sequences $\mathcal{Z}=(\mathcal{T}\cup\{\epsilon\})^{N}$,
where $\epsilon$ is a padding symbol, together with the map
$f_{\mathrm{rm-blanks}}:\mathcal{Z}\to\mathcal{X}$ that deletes all
padding. Given endpoints $x_0,x_1$ and a chosen alignment $z_0,z_1\in\mathcal{Z}$
with $f_{\mathrm{rm-blanks}}(z_j)=x_j$, the mixture path is defined on
the augmented space $\mathcal{X}\times\mathcal{Z}$ as

\begin{equation}
  p_t(x, z \mid z_0, z_1)
  = p_t(z \mid z_0, z_1)\,\delta_{f_{\mathrm{rm}\text{-}\epsilon}(z)}(x),
  \label{eq:editflow-aug}
\end{equation}

where $p_t(z\mid z_0,z_1)$ is the token-wise path applied on
the aligned sequences. The marginal rate of the CTMC that induces
this mixture path can be parametrized by a neural network that uses
a Bregman divergence objective~\citep{holderrieth2025generator,
havasi2025edit} during training.

\subsection{MAELLE} \label{app:maelle}

We take this section to detail the framework of MAELLE~\citep{maelle} a
discrete flow matching (DFM) model that performs pseudo-mechanism
- based forward reaction prediction. Discrete flow matching models
parametrize a Continuous Time Markov Chain (CTMC) that transports a
source distribution to a target distribution.

MAELLE treats molecules as augmented graphs $\mathcal{G} =
(\mathcal{V}, \mathcal{A}, \mathcal{E}, o)$, where $\mathcal{V}$
is the set of heavy atoms, $\mathcal{A}$ is the heavy-atom adjacency,
and $\mathcal{E}$ is a fixed set of \emph{electron sites} comprising
bonding sites $e_{ij}$ ($i \neq j$), lone-pair sites $e_{ii}$,
and per-atom hydrogen sites $e^H_i$ (as hydrogens on the same
heavy-atom are topologically equivalent, they are pooled into
a single site). 
The occupation vector $o \in \mathbb{Z}^{|\mathcal{E}|}$ counts electron
pairs at each site, including ones with zero occupation. In MAELLE's
framework, a forward reaction can be viewed as a heavy-atom conservative process
(\textbf{heavy}-atom conservative, since electron pairs
can be added at will on the per-atom hydrogen sites $e^H_i$) during which
the occupation vector $o$ is changed from the reactants to
in the products. The adjacency $\mathcal{A}$ can be recovered deterministically
from the current occupation vector.

The model performs modifications on the occupation vector by applying
\textbf{FLOW} (electron pair transferred from one electron site of the graph to
another) \textbf{ADD} (electron pair added to the molecular graph from a virtual electron source)
and \textbf{DEL} (electron pair removed from the molecular graph to a virtual electron sink)
\textit{electron edit} operations to it. These operations are defined as:

\begin{equation}
\begin{gathered}
    \text{\textbf{FLOW}}(e_{\text{source}} \rightarrow e_{\text{sink}})(o) = o - \mathbf{1}_{e_{\text{source}}} + 
    \mathbf{1}_{e_{\text{sink}}}, \text{requiring} \; o^{e_\text{source}}>0, \\
    \text{\textbf{ADD}}(e)(o) = o + \mathbf{1}_{e}, \\
    \text{\textbf{DEL}}(e)(o) = o - \mathbf{1}_{e}, \text{requiring} \; o^e>0,
\end{gathered}
    \label{electron_edits}
\end{equation}

where $\mathbf{1}_{e} \in \{0,1\}^{|\mathcal{E}|}$ denotes the one-hot
vector with a single nonzero entry at the position corresponding to
site $e \in \mathcal{E}$. Sequentially applying the electron edits
to the reactant molecular graph yields a pseudo-mechanistic trajectory.

The central object of a DFM model such as MAELLE is the conditional mixture
path that interpolates between the source distribution (in MAELLE's case,
reactants) and target distribution (in MAELLE's case, products). Given the
set of moves $\mathcal{M}$ that converts reactant electron occupation $o_0$
to product occupation $o_1$, denoted as $o_1 = \mathrm{ApplyEdit}
(o_0, \mathcal{M})$, and the subset $\mathcal{S}$ of moves are applied
in order to obtain the interpolated electron occupation $o$, MAELLE adopts
the edit-based mixture path:

\begin{equation}
  p_t(o \mid o_0, o_1)
  =
  \sum_{\mathcal{S} \subseteq \mathcal{M}}
  \underbrace{
  \kappa(t)^{|\mathcal{S}|}\,\bigl(1-\kappa(t)\bigr)^{|\mathcal{M}|-|\mathcal{S}|}
  }
  _{\displaystyle p_t(\mathcal{S}\,\mid\, o_0,o_1)}
  \;\mathbbm{1}\!\left[\,\mathrm{ApplyEdit}(o_0, \mathcal{S}) = o\,\right],
  \label{eq:edit-mixture}
\end{equation}

where $\kappa:[0,1]\to[0,1]$ is a monotone schedule with $\kappa(0)=0$ and $\kappa(1)=1$,
each move being applied with probability $\kappa(t)$. Thus, MAELLE's training target
becomes the marginal transition rate $u_t(m\mid o_t, \mathcal{G}_x)$ (where $m$ is
an electron edit operation) which characterizes the CTMC that
induces the edit-based mixture path described in Equation~\ref{eq:edit-mixture}.
This target is approximated by a graph neural network, $u_t^\theta$, which employs
for training a Bregman divergence objective~\citep{holderrieth2025generator,
havasi2025edit}. As for how the electron edits needed during training are obtained,
MAELLE solves a balanced integer Optimal Transport problem between $o_0$ and $o_1$,
using a simplex algorithm. The cost function between two electron sites is taken as the
minimum shortest-path distance between their constituent atoms on either the reactant
or product graph, while in the case of virtual sites used for DEL and ADD edits, a large,
constant penalty is used.

During inference, only the reactant is known. Thus, MAELLE initializes
$\mathcal{G}_t = \mathcal{G}_0$ at $t=0$ and evolves a CTMC trajectory
in $\left[0,1\right]$ for $N$ Euler steps, with a timestep of $\Delta t = \frac{1}{N}$. At each
step of the trajectory, every possible move $m$ fires independently
as a thinned Poisson process, having the following probability
of being applied:
\begin{equation}
  p_{\mathrm{fire}}(m)
  \;=\;
  1 - \exp\bigl(-\,u^\theta_t(m\mid\mathcal{G}_0,\mathcal{G}_t,t)\,\Delta t\bigr).
  \label{eq:fire}
\end{equation}
The electron occupation over the graph at the
end of the trajectory represents the reaction product.
As firing in Equation~\ref{eq:fire} is stochastic, running
multiple independent CTMC trajectories over the same reactant
will give a distribution of products. The frequency of each
product in the distribution gives the ranking for the
\textit{top-k} accuracy.

\section{Data pre-processing: USPTO 480K Synthon} \label{data_synthons}

The pipeline described in what follows is applied to all the entries of the training, validation and test partitions of the
USPTO-480K dataset~\citep{Lowe2017,jin2017predicting}, giving what we call the \textit{USPTO 480K Synthon} dataset.

Thus, a reaction is re-mapped using the RxnMapper
tool~\citep{schwaller2021extraction}.
Based on the atom mapping, reagents are then eliminated, leaving only the reactants and products.
Again based on the mapping, the heavy atoms that will form the synthons and the heavy atoms that will form the leaving
groups are identified: heavy atoms that are present both in the product and the reactants form the synthons, while heavy atoms
that are present in the reactants but not in the products form the leaving groups. The reactant SMILES are converted
into an RDKit molecule object and the bonds are iterated upon in the order given by the atom indices. If one of the
atoms within the bond is a synthon atom, while the other is a leaving group atom, the following rules are
applied in order to form synthons and leaving groups:
\begin{itemize}
    \item $\chi_{i} > \chi_{j}$ : bond between atom $\textit{i}$ and $\textit{j}$ is broken,
    atom $\textit{i}$ receives the charge $-BO$ and atom $\textit{j}$
    receives the charge $+BO$;
    \item $\chi_{i} \leq \chi_{j}$ : bond between atom $\textit{i}$ and $\textit{j}$ is broken,
    atom $\textit{i}$ receives the charge $+BO$ and atom $\textit{j}$ receives
    the charge $-BO$;
\end{itemize}
where \textit{i} and \textit{j} are atom indices with $i<j$ and $BO$ is the bond order of the bond being broken.
It is to note that the second condition is a design choice; putting the equality case to the first condition
would have worked equally well. Apart from the given products and reactants, the extracted synthons and leaving groups,
the atom maps of the indices between which bonds are broken are also saved in the dataset. Moreover, using the
products as source distribution and the synthons as target distribution, a pseudo-mechanistic
trajectory is interpolated between products and synthons using MAELLE's optimal transport pipeline~\citep{maelle}.

Finally, after deduplication based on the canonical SMILES string of the synthons, the \textit{USPTO 480K Synthon}
dataset is obtained, its characteristics being presented in Table~\ref{tab:synthon_dataset_stats}

\begin{table}[h]
    \centering
    \caption{Characteristics of the USPTO 480K Synthon dataset}
    \centering
    \small
    \setlength{\tabcolsep}{5pt}
    \begin{tabular}{@{}lccc@{}}
    \toprule
    \textbf{Partition} 
    & \textbf{Entries}
    & \textbf{Ring breaking}\\

    \textbf{} 
    & \textbf{}
    & \textbf{entries} [\%] \\
    \midrule
    Train & 384280 & 3.20\\
    Validation & 29678 & 3.23\\
    Test & 39461 & 3.12\\
    \bottomrule
    \end{tabular}
    
    \vspace{4pt}
    \label{tab:synthon_dataset_stats}
    \footnotesize{For a \textit{ring-breaking} entry, at least one ring is
    broken between the products and synthons}
\end{table}

\section{Training and inference details} \label{training_details}

We train Verso MAELLE on the USPTO 480K Synthon dataset
for product-to-synthon prediction using
the standard MAELLE architecture~\citep{maelle}, comprising of
approximately 16M trainable parameters. The AdamW
optimizer is employed, with $\beta_1 = 0.9$, $\beta_2 = 0.999$, initial
learning rate of $3\cdot10^{-4}$, weight decay of 0.01 and
ReduceLROnPlateau scheduler from PyTorch. We use an effective batch
size of 64 and train for 64 epochs (approximately 55 hours on a
single NVIDIA H100 GPU).

For inference with the base Verso MAELLE, we use 64 samples,
16 Euler steps, a Poisson temperature of 0.6, with sampling
for target electron site selection and an effective batch size of
4. For the FK steered Verso MAELLE with \textit{Ring Breaking Reward},
we use 16 samples, 16 Euler steps, a Poisson temperature of 0.6, 4
particles, $\lambda$ of 1, a sampling frequency of 4, with sampling for target electron site
selection and an effective batch size of 4.

For the case study of Punctaporonin U, we use for the FK-steered Verso MAELLE model the \textit{Domino
Reward}, with coefficients $c_{\textrm{ring}} = 1$, $c_{\textrm{flow}} = 0.2$ and $c_{\textrm{charge}} =1$. Furthermore, we use 32 samples, 16 Euler steps,
a Poisson temperature of 0.6, with sampling for target electron site selection,
16 particles, $\lambda = 5$ and a sampling frequency of 4. As for the predictions
on USPTO 480K Synthon, we employ the SUM potential. For the non-steered Verso MAELLE,
we use 32 samples, 16 Euler steps,
a Poisson temperature of 0.6, with sampling for target electron site selection.

For training and inference of the G2S, the configurations presented
in the original publication for forward-reaction prediction with D-GAT
on the mixed USPTO 480K dataset are used~\citep{tu2022permutation}. For the Molecular Transformer trained on
synthon prediction and for the
Synthon Transformer, we employ the configurations presented in the original
publication of the Molecular Transformer for forward-reaction prediction on the mixed USPTO 480K dataset~\citep{schwaller2019molecular}.
For the forward-reaction prediction Molecular Transformer,
we use the available weights of the model
trained on USPTO 480K mixed~\citep{schwaller2019molecular}.

For all the metrics that we report, we use the conventions
established by~\citet{maziarz2025re}, filtering out predicted
invalid synthons and deduplicating predictions based on
the canonical SMILES string (which has an influence over
the metrics of models that output SMILES autoregressively,
such as G2S and MT).

\section{Synthon Transformer}
\label{app:synthon_transformer}
\subsection{Synthon completion data format} \label{app:synthon_completion_data}

After the extraction of synthons from reactions, the
SMILES of the leaving groups may be enhanced through the addition of dummy atoms,
marked as $\textbf{[*]}$, to the
atoms that were previously bonded to synthon atoms. Furthermore,
the bond order to the dummy atoms is equal to the order of the bonds that were
broken. Finally, atom map numbers are added to the dummy atoms, equal to the current indices
(in the synthon SMILES string) of the
synthon atoms to which the leaving groups were previously bonded.
We thus obtain
a SMILES string representation of the leaving groups which can be unambiguously reattached
to the synthons in order to obtain the initial reactants. We call this representation
of the leaving groups \textit{enhanced SMILES}. We note that for synthons that
do not need leaving group completions, such as the reduced or oxidized forms of
the target products, the corresponding enhanced SMILES of the leaving groups
is given by the symbol \textbf{\$}.

Moreover, as the enhanced SMILES of the leaving groups are dependent
on the ordering within the SMILES string of the synthon, with each different ordering
producing different indices used for attachment, the data can be augmented
with random permutations of the synthon SMILES string. Such augmentations
have been shown to improve the performance of the Molecular Transformer~\citep{schwaller2019molecular}
and should therefore also improve the performance of the Synthon Transformer.

Thus, after synthons and leaving groups are extracted from USPTO 480K, we perform
one round of augmentation to the synthon SMILES and then convert the SMILES of the
leaving groups to the corresponding enhanced SMILES string.

\subsection{Leaving group prediction by Synthon Transformer} \label{app:transformer_pred}

For the Synthon Transformer, we report here the \textit{top-k leaving group
prediction} that was obtained on the test set of USPTO 480K Synthon. We note that
the input for the model for this metric was the ground truth synthons of Synthon
USPTO 480K, and not the synthons predicted by the models. Thus, this metric
measures strictly the capacity of the Synthon Transformer to accurately predict
leaving groups.

\begin{table}[ht]
    \centering
    \caption{Leaving group prediction on USPTO 480K Synthon dataset}
    \small
    \setlength{\tabcolsep}{3pt}
    \label{tab:topk-lg}
    \begin{tabular}{@{}lccccc@{}}
    \toprule
    & \multicolumn{5}{c}{\textbf{Top-\textit{k} Leaving Group Accuracy}} \\
    \cmidrule(lr){2-6} 
    \textbf{Models} & $k=1$ & $k=2$ & $k=3$ & $k=4$ & $k=5$  \\
    \midrule
    Synthon Transformer              & 0.7655 & 0.8678 & 0.8966 & 0.9087 & 0.9131 \\
    \bottomrule
    \end{tabular}
    
    \vspace{4pt}
    \footnotesize{The Synthon Transformer was trained on
     USPTO 480K Synthon with one round of
    augmentation}
\end{table}


\newpage

\end{document}